\documentclass[aps,prb,twocolumn,superscriptaddress]{revtex4-1}
\usepackage{subfigure}
\usepackage{graphicx}
\usepackage{natbib}
\usepackage{amsmath}
\usepackage{amssymb}

\begin{document}

\title{Capturing long range correlations in two-dimensional quantum lattice systems using correlator product states}


\author{S. Al-Assam}
\email{s.al-assam1@physics.ox.ac.uk} \affiliation{Clarendon Laboratory, Department of
Physics, University of Oxford, Parks Road, Oxford, OX1 3PU, United Kingdom}
\author{S. R. Clark}
\affiliation{Centre for Quantum Technologies, National University of Singapore, 3 Science
Drive 2, Singapore 117543} \affiliation{Clarendon Laboratory, Department of Physics,
University of Oxford, Parks Road, Oxford, OX1 3PU, United Kingdom}
\author{C. J. Foot}
\affiliation{Clarendon Laboratory, Department of Physics, University of Oxford, Parks
Road, Oxford, OX1 3PU, United Kingdom}
\author{D. Jaksch}
\affiliation{Clarendon Laboratory, Department of Physics, University of Oxford, Parks
Road, Oxford, OX1 3PU, United Kingdom} \affiliation{Centre for Quantum Technologies,
National University of Singapore, 3 Science Drive 2, Singapore 117543}

\date{\today}

\begin{abstract}
We study the suitability of correlator product states for describing ground-state
properties of two-dimensional spin models. Our ansatz for the many-body wave function
takes the form of either plaquette or bond correlator product states and the energy is
optimized by varying the correlators using Monte Carlo minimization. For the Ising model
we find that plaquette correlators are best for estimating the energy while bond
correlators capture the expected long-range correlations and critical behavior of the
system more faithfully. For the antiferromagnetic Heisenberg model, however, plaquettes
outperform bond correlators at describing both local and long-range correlations because
of the substantially larger number of local parameters they contain. These observations
have quantitative implications for the application of correlator product states to other
more complex systems, and they give important heuristic insights: in particular the
necessity of carefully tailoring the choice of correlators to the system considered, and
its interactions and symmetries.
\end{abstract}

%\keywords{}

\maketitle
\section{Introduction}
\subsection{Background}
Modeling strongly correlated systems exactly for more than a few particles is not
possible due to the rapid increase in the size of the Hilbert space with their number.
However, there has been much success recently in using tensor network methods to
numerically simulate strongly correlated systems.\cite{Cirac:2009p5944} These approaches
allow states with suitable properties, such as obeying an area law for their entanglement
entropy,\cite{arealaw1,*arealaw2} to be efficiently represented by a network of tensors.
They also provide efficient methods of `contracting' the network to allow expectation
values of operators, or more basically wave function amplitudes, to be calculated. In
particular, matrix product states (MPS) can be used to accurately describe large systems
in one dimension and can be exactly and efficiently
contracted.\cite{Verstraete:2006p5986} Crucially this means that expectation values can
be calculated in a number of steps that only grows as a low degree polynomial with system
size and tensor size. This has been crucial for the success of MPS based algorithms like
the density-matrix renormalization
group,\cite{WhitePRL,*WhitePRB,*Schollwock:2005p3990,*Schollwock:2010p3876,*DMRGPlenio}
and time-evolving block decimation.\cite{Vidal2003,*Vidal:2004p5981,*Clark2004} Tensor
networks can be extended to higher dimensions using the projected entangled pair states
(PEPS) construction,\cite{Verstraete:2004,*Murg:2007p3742} but unlike MPS these do not
permit efficient exact contraction. Approximate contraction procedures in two dimensions,
such as those based on MPS methods,\cite{Verstraete:2008p3835,Cirac:2009p5944} or tensor
renormalisation group (TRG),\cite{TRG1, TRG2, TRG3} must be employed at the cost of
having a much less favorable, though polynomial, scaling. Recently a procedure for
performing PEPS simulations combining TRG and Monte Carlo sampling has been proposed, and
it shows promising results for treating large bond dimensions.\cite{Wang:2010p7758} In
contrast to the PEPS construction, both the multiscale entanglement renormalisation
ansatz (MERA)\cite{MERA1,*MERA2} and tensor tree networks
(TTNs)\cite{Tagliacozzo:2009p3620} utilize a hierarchical structure where the tensors are
interconnected by bonds according to a tree pattern, and can be contracted exactly and
efficiently. Like PEPS, these methods scale as a high degree polynomial.

An alternative less computationally expensive approach to represent two-dimensional
systems is to consider classes of states that while not being efficiently contractible
exactly, are efficiently and exactly samplable, i.e.\ for any given configuration the
amplitude can be found exactly. The tensor network formalism provides a powerful
framework from which to devise new and physically tailored ans\"{a}tze with these
properties for which the closest approximate ground state can be found using variational
Monte Carlo methods. For example, string bond states use a product of overlapping MPS
`strings' and acquire their exact samplability from the contractibility of the underlying
MPS.\cite{Schuch2008sbs} An alternative ansatz is formed by building the wave function
using superpositions of all possible coverings of singlets i.e.\ resonating valence bond
(RVB) states\cite{RVBHeis, SandvikVB, lousvk2007,SandvikEvertz}, and it has been shown
that RVB states have a PEPS description.\cite{PEPSRVB} Recently a new ansatz within this
class has been proposed for simulating lattice systems: so-called correlator product
states, \cite{changlani2009,Neuscamman:2010p7518} or entangled plaquette
states,\cite{Mezzacapo2009,Mezzacapo:2010p4750,PhysRevB.83.115111,
MezzacapoJ1J2Honeycomb} which are equivalent and are hereafter referred to as CPS. A
similar approach was also proposed in some earlier
works.\cite{NightingaleBlote,*HuseElser,*Nishino1,*Nishino2} Correlator product states
can be thought of as a more basic form of tensor network states, where correlations
between sites are encoded explicitly in ``correlator'' building blocks so that the
amplitude for each configuration is given by a product of their scalar elements. They
provide a slightly simpler ansatz than string bond states, but with similar power and
properties.\cite{changlani2009}

\subsection{Motivation}

The CPS construction has been applied to a variety of models, and in all cases the
energies found compared well with those found using other methods, e.g.\ PEPS, MPS,
stochastic series expansion (SSE), but with a much reduced computational cost. As we will
discuss in detail below, the CPS description is incredibly versatile, and the correlator
type can be varied without adversely affecting the complexity of the calculation. Thus,
the type of correlator can be chosen to best match the properties or symmetries of the
system. It is also possible to go beyond PEPS, which are composed of short ranged bonds,
and consider CPS that contain long ranged bonds and that are area-law
violating.\cite{arealaw1,*arealaw2}

In previous work, the estimated ground state energy as a function of the plaquette size
has been determined.\cite{Mezzacapo2009} What has so far been lacking is a systematic
study of the relative behavior of different correlator types for describing the important
physical properties of different systems. In particular bond correlators, where
correlators are arranged to connect pairs of sites across the lattice, have not been
studied in any detail. Bond correlators allow the possibility to directly encode
long-range correlations between distant sites using a very small number of parameters.
This is not possible with plaquette correlators, due to the exponential increase in the
number of correlator elements with plaquette size, or with matrix product states, where
distant correlations are mediated by intermediate nearest neighbor bonds. However,
plaquette correlators do provide a more straightforward ansatz that imposes a less rigid
short ranged structure. It is thus interesting to investigate what effect the choice of
correlator product ansatz has on the ability to describe systems that possess long-range
correlations, as well as the difficulties they may present during minimization.

The CPS construction seems promising for describing a wide variety of spin models, and
modeling these systems could provide insights for cold-atom simulations of quantum
magnetism.\cite{QMag1,QMag2} In previous works it has also been noted that certain
important states such as the Laughlin wave function,\cite{LaughlinPRL} which cannot be
efficiently described with other types of tensor network states,\cite{IblisdirPRL} can
have an exact CPS description when bond correlators between all sites are
included.\cite{changlani2009} This illustrates that bond correlators can efficiently
describe complex topological systems, and may be able to describe other fractional
quantum Hall states.\cite{QHE-ezawa} To aid future studies which will apply the CPS
ansatz to more complicated systems, in this work we test their effectiveness using
systems whose behavior is well-known. We examine in detail the performance of different
CPS ans\"{a}tze, but note that there is no guarantee that the state with the lowest
estimate of the energy better reproduces any properties of the ground state, aside from
of course energy, better than another with a higher energy estimate.\cite{Messiah}
Therefore, as well as estimating the ground state energy, we also investigate the
critical behavior, the long-range correlations and antiferromagnetic order, which provide
useful benchmarks of their effectiveness. Our observations provide insight of
quantitative and heuristic value into the accuracy and applicability of the CPS approach.

\subsection{Outline and main results}
In this paper, we investigate different correlator types, with the aim of identifying
those that are most effective at capturing the behavior of the system, and find that the
optimum correlator type is strongly dependent on the physical properties of the system.
In section \ref{sec:methods}, we describe the different correlators types and outline the
methods used to determine the ground state properties. In section \ref{sec:isres} we
examine the effectiveness of the different correlator types at describing the long range
order and critical phenomena by applying them to the quantum transverse Ising model (TIM)
on a square lattice. We find that bond correlators are better able to predict the
critical point and represent expected long range correlations than plaquette correlators
for the two-dimensional TIM, where bond correlators allow a larger range of sites to be
covered for a given number of parameters and computational effort. In section
\ref{sec:istrihex} we extend the treatment to different lattice geometries, finding
similar behavior for the performance of the two correlator types. In section
\ref{sec:heis} we investigate the ability of plaquette and bond correlators to describe a
more complex type of long range order and antiferromagnetism by applying them to the
antiferromagnetic Heisenberg model (AFHM). We find that plaquette correlators are more
successful at describing the AFHM, since they provide a much larger number of local
parameters to better capture the complicated local structure encoding the
antiferromagnetic order in this system. Finally, we conclude and summarize the results in
section \ref{sec:conc}.

\section{Method} \label{sec:methods}
The exact and efficient samplability of CPS makes them ideal for variational minimization
(for details of the exact-efficient samplability of CPS see appendix \ref{ap:samp}).
Previously CPS have been minimized using a generalized eigenvalue method,
\cite{changlani2009} and a deterministic method.\cite{Neuscamman:2010p7518} Here we use a
Monte Carlo based stochastic minimization method to determine the ground
state,\cite{lousvk2007} which has previously been successfully applied to one-dimensional
systems represented by matrix product states.\cite{SandvickVidal2007} The correlator
elements that best approximate the ground state are found by estimating the derivative of
the energy with respect to each correlator element, and then updating each correlator
element according to the direction of the derivative with a random step size. This allows
minimization using only the first derivative of the energy (so requiring fewer
computation steps), and ameliorates the error associated with updating the correlator
elements. Precise details of this numerical method are described in appendix
\ref{ap:meths}.

\subsection{System set-up}
We consider a system with $N$ sites where each site $i$ has an identical local Hilbert
space spanned by states $|s_i\rangle$. The full Hilbert space of the system is then
spanned by states $|\mathbf{s}\rangle = |s_1\rangle \otimes |s_2\rangle \otimes \cdots$
where we term $\mathbf{s} = (s_1, s_2, \cdots)$ the configuration of the state. The
many-body wave function is $|\psi\rangle = \sum_\mathbf{s} W(\mathbf{s})
|\mathbf{s}\rangle$, where $W(\mathbf{s})$ is the amplitude or weight of a given
configuration $\mathbf{s}$. In this work we will for simplicity consider only
spin-$\frac{1}{2}$ systems, but the methods used can be applied straightforwardly to
lattice systems possessing a larger on-site dimension.

\begin{figure}[t]
\begin{center}
\includegraphics[scale = 0.55]{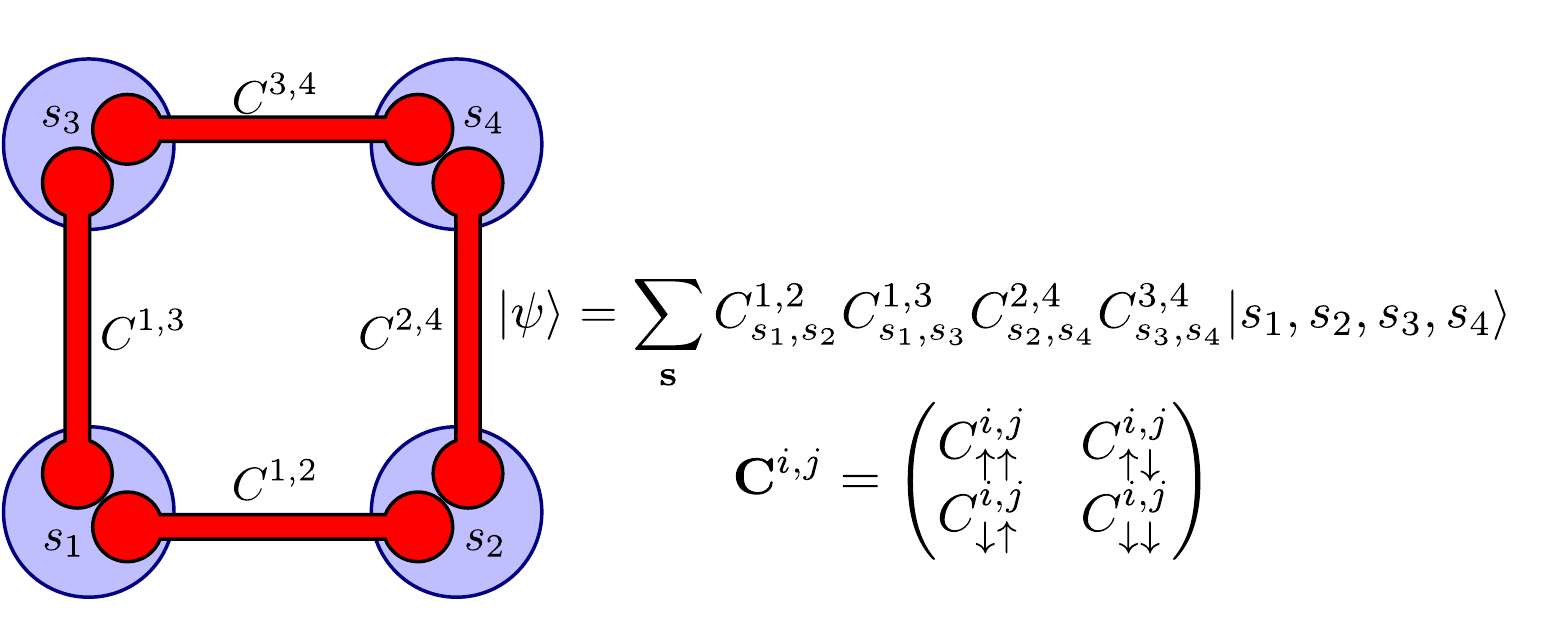}
\caption{(Color online) Illustration of the CPS ansatz for a state $|\psi\rangle$
describing a simple $2 \times 2$ system with open boundary conditions. The local states
$s_{i}$ are those of a spin-$\frac{1}{2}$ system. Each correlator $\mathbf{C}^{i,j}$
represents the amplitudes for different configurations of spins on two sites $i,j$. The
four-site system has a total of sixteen basis states, with the weight for an example
basis state \mbox{$W(\uparrow\downarrow\downarrow\uparrow) =
C^{1,2}_{\uparrow\downarrow}C^{1,3}_{\uparrow\downarrow}C^{2,4}_{\downarrow\uparrow}C^{3,4}_{\downarrow\uparrow}$}.}
\label{fig:cpsanstaz}
\end{center}
\end{figure}

In the correlator product state description, the weight $W(\mathbf{s})$ is given by the
product of correlator elements $C^{\{i\}}_{s_{\{i\}}}$ over the lattice
\begin{equation}
W(\mathbf{s}) = \prod_{\{i\}}C^{\{i\}}_{s_{\{i\}}},
\end{equation}
where each correlator element is a $c$-number that describes the amplitude of a
configuration $s_{\{i\}}$ of a subgroup of sites $\{i\}$.  The wave function in the CPS
representation is given by
\begin{equation}
|\psi\rangle = \sum_{\{s\}} \prod_{\{i\}}C^{\{i\}}_{s_{\{i\}}}|s_{1}\cdots s_{L}\rangle.
\end{equation}
We illustrate this in Fig.\ \ref{fig:cpsanstaz}, for the case where the subgroup of sites
is a nearest neighbor pair i.e.\ `bond' correlators, for a $2 \times 2$ system.

This compact description of a state can be extended to any form of correlator, for
example each correlator could represent four sites in a plaquette, or a string of a given
number of sites. One advantage of using CPS is that the wave function amplitude is a
simple product of $c$-numbers. Unlike with PEPS, the amplitude $W(\mathbf{s})$ can be
calculated efficiently and exactly for any configuration $\mathbf{s}$. This allows the
wave function to be efficiently sampled, so that Monte Carlo methods can be used to
determine expectation values and minimize the wave function. The flexibility and
exact-efficient samplability of CPS is described further in appendix \ref{ap:samp}.

\subsection{Correlator types} \label{sec:corrtypes}
In the following, we study the ground states using different types of correlator, and
compare their properties. We use periodic boundary conditions with the translational
invariance of the system allowing one correlator of each type to describe the entire
system.

\begin{figure}[t]
\begin{center}
  \includegraphics{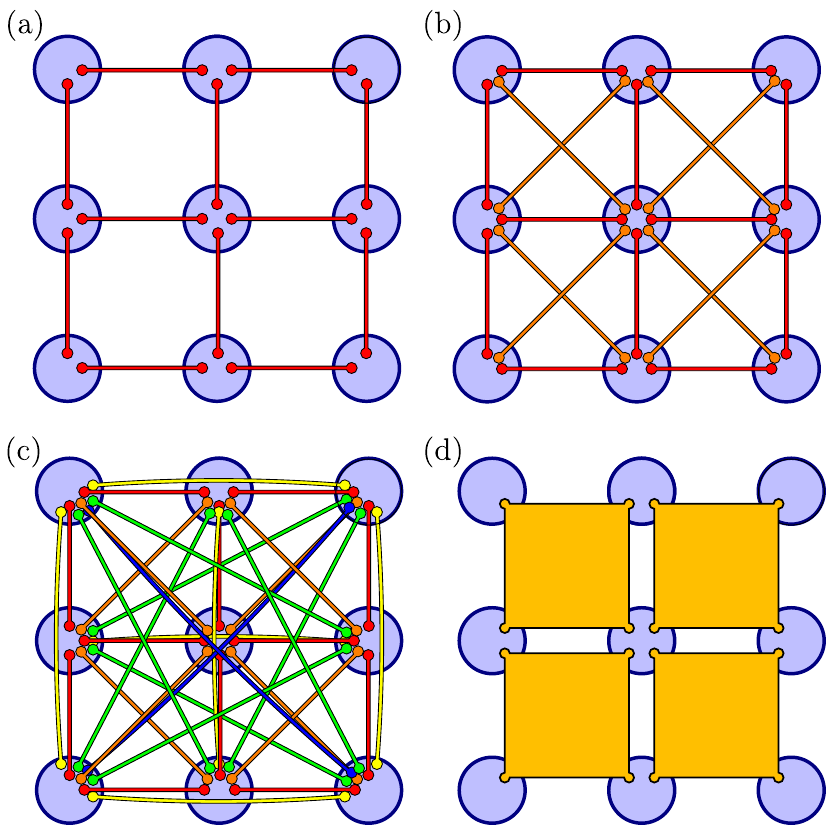}
  \caption{(Color online) Illustration of the different correlator types used. The blue circles represent
  lattice sites, and only correlators that link these $3 \times 3$ sites are represented. (a) Nearest neighbor (n.\ n.) bond correlators.
  (b) Bond correlators with $r_{\mathrm{max}} = 1$. (c) Bond correlators with $r_{\mathrm{max}} =
  2$. (d) $2 \times 2$ plaquette correlators.}
  \label{fig:cpstypes}
\end{center}
\end{figure}

The types of correlators used for square lattices are illustrated in Fig.\
\ref{fig:cpstypes}. Nearest-neighbor (n.\ n.) bond correlators, with one correlator for
the vertical bonds and one correlator for the horizontal bonds as shown in Fig.\
\ref{fig:cpstypes}(a), are the simplest correlator types used. This gives a description
of the ground state using only eight parameters. We also use bond correlators with longer
range bonds. In general we term a bond correlator of size $r_{\mathrm{max}}$ as including
all bond vectors up to the maximum range $\Delta\mathbf{r}_{\mathrm{max}} =
(r_{\mathrm{max}},r_{\mathrm{max}})$ i.e.\ it includes all the diagonal bonds as well as
bonds along the lattice axes. Figures \ref{fig:cpstypes}(b) and (c) show the set-up with
correlators of size $r_{\mathrm{max}} = 1$ and $r_{\mathrm{max}} = 2$ respectively, and
in the following calculations we use correlators up to $r_{\mathrm{max}} = 8$.  For
correlators of size $r_{\mathrm{max}}$, the number of correlator elements is
$8r_{\mathrm{max}}(r_{\mathrm{max}}+1)$. To calculate the energy, the number of
computational steps scales as $O(N^2r_{\mathrm{max}}^2)$, and the only part of the
calculation that depends on the correlator size and type is the calculation the
correlator fraction given in equation (\ref{eq:prob}).

We also use plaquette correlators, with the smallest $2 \times 2$ plaquette illustrated
in Fig.\ \ref{fig:cpstypes}(d), up to a size of $4 \times 4$ plaquettes. The correlators
are set up so that they are displaced from one another by one site and overlap one
another. In general, the greater the overlap between the correlators the more accurate
the description of the ground state. The number of elements in an $n \times n$ correlator
scales as $2^{n^{2}}$, so the memory requirements for a given plaquette correlator size
scale much faster than for bond correlators. However the number of computational steps
required to calculate the energy is $O(N^2n^4)$, which is a mild polynomial scaling and
grows much more slowly than the number of correlator elements. The reason for this is
that calculating an off-diagonal expectation value only involves picking out the correct
correlator element for the subgroup of sites spanned by a correlator. As a result it
depends on the number of sites spanned by the correlator and the number of correlators
that fall on a given site, and not on the number of elements. Note that calculation of
the energy derivative formally requires a number of steps that does scale with the number
of correlator elements, however in practice this part of the calculation is fast compared
to estimating the energy itself and it is not rate limiting. Thus plaquette correlators
allow the use of many more parameters for describing the system with only a modest
increase in computational effort, although in our calculations their size is ultimately
limited to $4 \times 4$ by memory requirements.

Comparing the computation time for correlator types, a calculation using $3 \times 3$
plaquette correlators has approximately the same number of computation steps as for bond
correlators up to $r_{\mathrm{max}} = 4$, and a calculation using $4 \times 4$ plaquette
correlators has approximately the same number of computation steps as for bond
correlators up to $r_{\mathrm{max}} = 8$. However, when comparing the performance of
different correlator types, it is worth considering the three main differences between
them: (i) The number of sites that can be reached from a given site using the correlator:
$r_{\mathrm{max}} = 1$ is equivalent to a $2 \times 2$ plaquette, $r_{\mathrm{max}} = 2$
is equivalent to a $3 \times 3$ plaquette, and $r_{\mathrm{max}} = 3$ is equivalent to a
$4 \times 4$ plaquette. (ii) The computational effort: for a given number of spanned
sites, our routines for bond correlators are 2, 3.5 and 5.5 times faster than for
plaquette correlators respectively. (iii) The number of correlator elements: the number
of correlator elements for a given number of sites spanned is far smaller for bond
correlators. In addition to the reduction in computational effort, this makes it possible
to span far more sites using bond correlators. The more fragmented structure of bond
correlators also permits different minimization strategies as described in more detail in
appendix \ref{ap:meths}.

\section{Long range correlations and critical behavior: the quantum transverse Ising model on a square lattice} \label{sec:isres}

As an ideal test case for examining different correlator types, we consider the TIM on an
$L \times L$ square lattice, described by the Hamiltonian
\begin{equation}
H =
-J\sum_{i=1,j=1}^{L}(\sigma^{[i,j]}_{z}\sigma^{[i+1,j]}_{z}+\sigma^{[i,j]}_{z}\sigma^{[i,j+1]}_{z}
+ g\sigma^{[i,j]}_{x}),
\end{equation}
where $J>0$ is the coupling, $i$ and $j$ denote the lattice index in the two
perpendicular directions, $\sigma^{[i,j]}_{z (x,y)}$ is the Pauli $z (x,y)$ operator for
spin $(i,j)$, and $g$ is the dimensionless transverse magnetic field. We consider this
model because it is one of the archetypal systems that exhibits a quantum phase
transition at a finite magnetic field.\cite{SachdevQPT} The system moves from an ordered
ferromagnet in the $z$ direction at $g=0$ to a state disordered in the $z$ direction and
aligned in the $x$ direction when $g\gg1$. As the system approaches criticality at
$g=g_c$, the correlation length diverges, and the system becomes gapless, making a
numerical description of the state challenging. The critical point in the thermodynamic
limit, found using a finite-size scaling analysis,\cite{Hamer:2000p3494} is $g_{c}
\approx 3.044$, where the pseudo-critical point\cite{Note1} was defined using a careful
extrapolation of the ratio of the energy gap between consecutive system
sizes.\cite{Barber}

By applying the CPS ansatz to this system, we investigate its performance at describing
the behavior of several important physical quantities as the transverse magnetic field is
varied. We focus in particular at modeling the long-range correlations close to the
pseudo-critical point, and compare with the many previous studies performed using other
numerical and (approximate) analytical methods. This system is thus a highly useful
benchmark of the method for demonstrating both the effectiveness and limitations of the
CPS approach.

\subsection{Energy} \label{sec:isingen}
\begin{table}
\begin{center}
\begin{tabular}{|c||c|r@{.}l|c|}
  \hline
  Correlator type & No. of elements &\multicolumn{2}{|c|}{Energy}  & Error \\
  \hline
  \hline
  n.\ n.\ bonds & 8 & $-3$ & $2348(4)$ & $13 \times 10^{-3}$ \\
  ${r}_{\mathrm{max}} = 1$ & 16 & $-3$ & $2384(3)$ & $9\times 10^{-3}$ \\
  ${r}_{\mathrm{max}} = 2$ & 48 & $-3$ & $2457(3)$ & $2\times 10^{-3}$ \\
  ${r}_{\mathrm{max}} = 3$ & 96 & $-3$ & $2465(2)$ & $1\times 10^{-3}$ \\
  \hline
  $2 \times 2$ plaquettes  & 16 & $-3$ & $2387(4)$ & $9 \times 10^{-3}$ \\
  $3 \times 3$ plaquettes & 512 & $-3$ & $2463(2)$ & $1\times 10^{-3}$ \\
  $4 \times 4$ plaquettes & 65,536 & $-3$ & $24725(5)$ & $2\times 10^{-5}$ \\
  \hline
\end{tabular}
\end{center}
\caption{Energy per site in units of $J$ found using stochastic minimization for the
two-dimensional TIM in a $6 \times 6$ system for different correlator types. Results are
compared with exact results at $g = 3.05266$, which has been calculated to be the
pseudo-critical point in a $6 \times 6$ system, with a ground state energy of
$-3.2472744\cdots$ (Ref.\ \onlinecite{Hamer:2000p3494}).} \label{tb:6sysy}
\end{table}

The energy and its derivative, as well as other observables, are calculated using the
algorithm described in appendix \ref{ap:meths}.  Note that for all correlator types we
restrict to only real parameters without any loss of generality, since for this model the
ground state can be constructed using real, positive weights for all configurations. We
first calculate the energy at a value of the magnetic field close to the critical point
for two systems. The largest system that has been solved numerically exactly is a $6
\times 6$ system,\cite{Hamer:2000p3494} and so provides a good comparison for how well
the method is working. This system size has also been solved using
TTNs,\cite{Tagliacozzo:2009p3620} so we also compare the accuracy of the CPS method to
those results. The energy of the ground state found using different correlator types is
shown in Table \ref{tb:6sysy}. The error, defined as the difference between the energy
estimated using CPS and the exact ground state energy, is also displayed.
%Exact energy is -3.247274397582149
As expected, we find that the energy converges to the exact value for increasing
correlator size. In comparison, for TTNs, around $10^7$ parameters are required to reduce
the error to $2\times 10^{-5}$.\cite{Tagliacozzo:2009p3620} We also apply the CPS method
to much larger systems. In Table \ref{tb:31sys} we show the minimized energy in a $31
\times 31$ system at $g = 3.05$, which is close to the pseudo-critical point.

\begin{table}
\begin{center}
\begin{tabular}{|c||c|c|}
  \hline
  Correlator type & No. of elements & Energy \\
  \hline
  \hline
  n.\ n.\ bonds & 8 & $-3.2325(1)$\\
  ${r}_{\mathrm{max}} = 1$ & 16 & $-3.2357(1)$\\
  ${r}_{\mathrm{max}} = 2$ & 48 & $-3.23815(8)$ \\
  ${r}_{\mathrm{max}} = 3$ & 96 & $-3.23865(5)$ \\
  ${r}_{\mathrm{max}} = 5$ & 240 & $-3.23866(7)$ \\
  ${r}_{\mathrm{max}} = 8$ & 576 & $-3.23882(7)$ \\
  \hline
  $2 \times 2$ plaquettes & 16 & $-3.23599(9)$ \\
  $3 \times 3$ plaquettes & 512 & $-3.23864(4)$ \\
  $4 \times 4$ plaquettes & 65,536 & $-3.23903(5)$ \\
  \hline
\end{tabular}
\end{center}
\caption{Energy per site in units of $J$ found using stochastic minimization for the
two-dimensional TIM in a $31 \times 31$ system for different correlator types at $g =
3.05$.} \label{tb:31sys}
\end{table}

We find that even though plaquette correlators have more elements, the energy convergence
is more well-behaved than for bond correlators. In line with commonly known properties of
non-convex optimization, it is likely that having a larger number of parameters is
beneficial when there is a foliated energy landscape since it allows the optimization
more freedom in finding the minimum energy state. Conversely constraining the number of
parameters for such long range bonds tends to result in numerous local minima giving a
more difficult minimization problem. For example in a previous work minimizing over a
similar class of many-body states, it was found that the enforcement of certain
symmetries increased the difficulty of finding a good state, since this amounted to
cutting through the energy landscape of the parameter space, dividing it into separated
minima.\cite{AndersWGS} However, even though a larger span of sites is found to be needed
to minimize to a given energy with bond correlators, the number of parameters required is
still far smaller than for plaquettes substantially reducing the computational effort.

\begin{figure}[t]
\begin{center}
\includegraphics{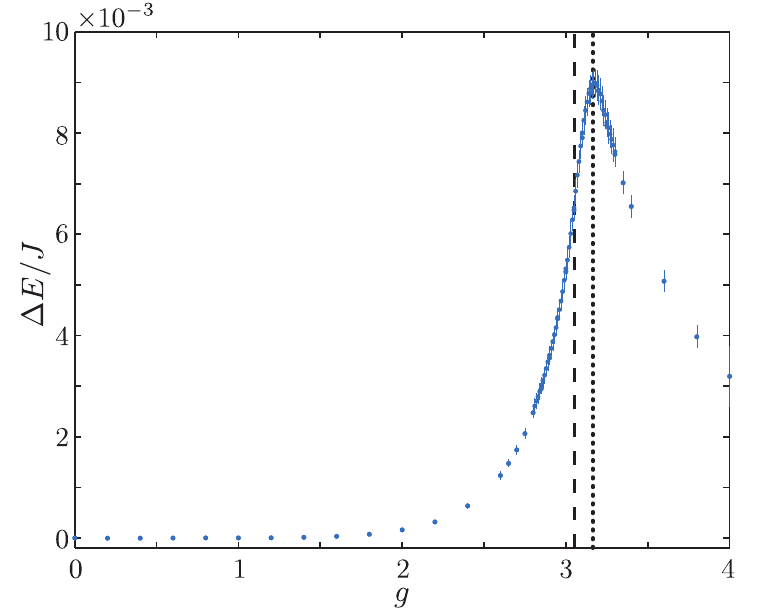}
\caption{(Color online) Difference in energy between ground state calculated using
nearest-neighbor bond correlators, and ground state calculated using $4 \times 4$
plaquette correlators.  The dashed line indicates the critical magnetic field found by a
finite-size scaling analysis (Ref.\ \onlinecite{Hamer:2000p3494}), and the dotted line
indicates the peak of the energy difference at $g = 3.17$ --- this is approximately
midway between the pseudo-critical points found with nearest-neighbor bond and $4 \times
4$ plaquette correlators.} \label{fig:ediff}
\end{center}
\end{figure}

We also investigate how the choice of correlator affects the energy at different values
of the transverse magnetic field $g$. Figure \ref{fig:ediff} shows the difference between
the energy calculated using n.\ n.\ bond correlators (which give the highest estimated
ground state energy) and the energy calculated using $4 \times 4$ plaquette correlators
(which give the lowest estimated ground state energy), for different values of the
transverse magnetic field $g$. There is not much difference between the two values for
$|g - g_{c}| \gg 1$, but the energy difference increases when $g \approx g_{c}$. Far from
criticality, when the correlations are expected to be short range, the energy can be
calculated accurately using a small number of parameters, however as the correlation
length increases a larger number of parameters are needed to describe the system
accurately. The maximum energy difference is still small ($< 1 \%$), however this can
lead to large differences in the properties of the state due to the large number of
low-lying excited states combined with a vanishing gap as $g_c$ is approached. This is a
general problem with using a variational approach to describe critical systems. The
investigation of these properties is described in the following sections, and we find
that choosing an ansatz with a suitable structure helps to better describe some important
physical properties of the ground state even close to criticality. Even though formally
the energy found using bond correlators is larger the structure of the ansatz seems to
favor those states that possess long-range correlations.

\subsection{Transition point}

\begin{figure}[t]
\begin{center}
\includegraphics{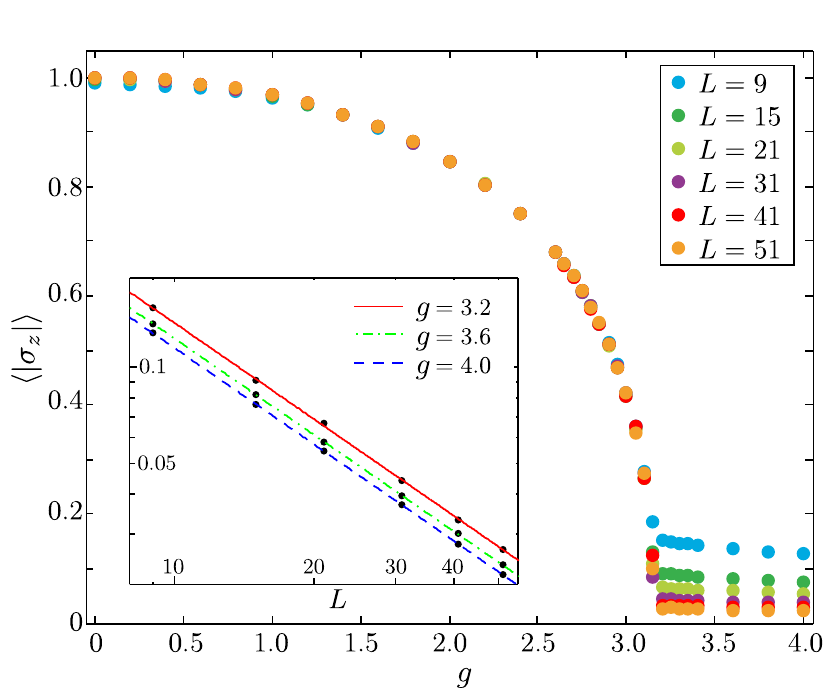}
\caption{(Color online) Absolute magnetization $\langle|\sigma_{z}|\rangle$ as a function
of transverse magnetic field $g$ in an $L \times L$ system, for the ground state found
using $2 \times 2$ plaquette correlators. The pseudo-critical point occurs at $g \approx
3.15 $. \textsc{Inset:} $\langle|\sigma_{z}|\rangle$ at constant $g>g_c$ as a function of
$L$, both axes having a logarithmic scale, with a fit of the form
$\langle|\sigma_{z}|\rangle \propto L^{-b}$. The data was fitted with $b = 1.0$.}
\label{fig:magvB1p}
\end{center}
\end{figure}

We investigate the position of the pseudo-critical point in the system by examining the
local order parameters. Specifically, we calculate both the transverse magnetization
$\langle\sigma_x\rangle$, and the absolute magnetization $\langle|\sigma_z|\rangle$,
defined as the expectation of the absolute value of the average of all $L \times L$
spins, i.e.\ it quantifies how well the spins are aligned with one another (N.B.\
$\langle\sigma_z\rangle$ is zero due to the global $\mathbb{Z}_2$ symmetry of the
system). Figure \ref{fig:magvB1p} shows the results for the absolute magnetization
$\langle|\sigma_z|\rangle$ as a function of transverse magnetic field using $2 \times 2$
plaquette correlators for a number of $L \times L$ system sizes. The error bars for each
point, where the error is given as the standard deviation of the $G$ different bins (see
appendix \ref{ap:meths}), are smaller than the marker for the data point.

As expected, we find a sharp drop in the magnetization at around $g = 3.05$ and the
change in magnetization becomes steeper as the system size increases. We also find that
the small non-zero magnetization for $g > g_{c}$ decreases as the system size increases:
for $L = 51$, the magnetization at $g = 4$ is $\langle |\sigma_{z}|\rangle \approx
0.022$. The inset in Fig.\ \ref{fig:magvB1p} shows the magnetization as a function of
lattice size for different fixed values of $g>g_c$, with a fit to a power law decay of
the magnetization, i.e.\ $\langle |\sigma_{z}|\rangle_{g>g_c} = aL^{-b}$. The data fit an
inverse scaling with system size, indicating that the magnetization decays to zero for
$g>g_c$ in an infinite system. While the behavior found for $2 \times 2$ correlators is
not quantitatively precise (the pseudo-critical point occurs at $g \approx 3.15$ rather
than at $g = 3.044$), it is simple and offers a computationally affordable means of
approximately locating a critical point once the appropriate local order parameters
signifying it are known.

\begin{figure}[t]
\begin{center}
\includegraphics{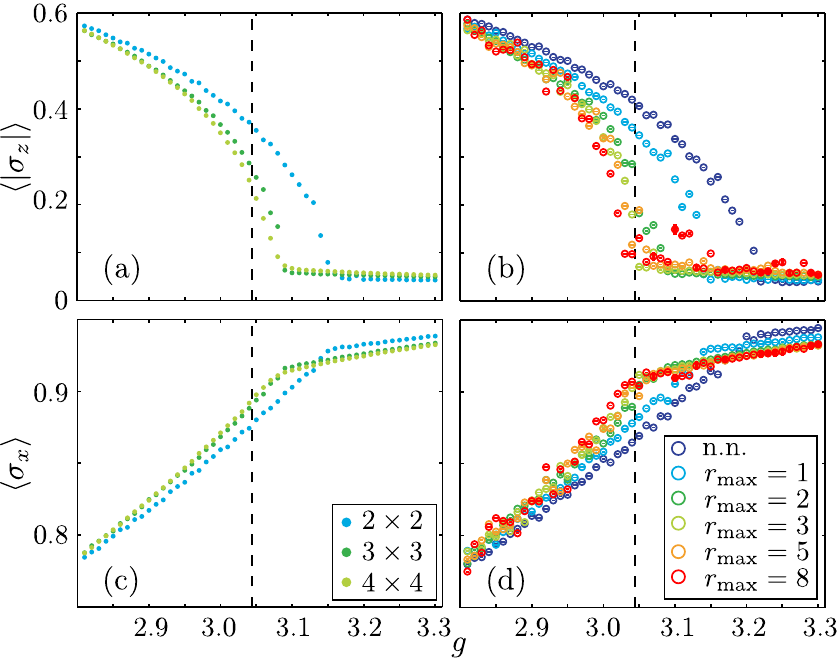}
\caption{(Color online) Absolute magnetization [(a) and (b)] and transverse magnetization
[(c) and (d)] as a function of transverse magnetic field in a $31 \times 31$ system, for
the ground state found using different sized plaquette correlators [(a) and (c)] and bond
correlators [(b) and (d)].  The dashed line indicates the critical magnetic field found
by a finite-size scaling analysis in Ref.\ \onlinecite{Hamer:2000p3494}, which occurs at
$g_c$ = 3.044.} \label{fig:mvBmp}
\end{center}
\end{figure}

As described in section \ref{sec:isingen}, we find that larger correlators are better
able to minimize the energy. Figure 5 shows how the behavior of the local order parameter
$\langle|\sigma_z|\rangle$ and the transverse magnetization $\langle\sigma_x\rangle$
depend on the correlator type and size. The position of $g_c$, the critical point found
by a finite-size scaling analysis,\cite{Hamer:2000p3494} is also indicated. The plots of
the absolute magnetization $\langle |\sigma_{z}|\rangle$ shown in Figures
\ref{fig:mvBmp}(a) and (b) show that as the correlator size increases, the calculated
pseudo-critical point moves closer to $g_c$ for both plaquette and bond correlators. The
transverse magnetization (shown in Figures \ref{fig:mvBmp}(c) and (d)) displays the
expected `knee' at the pseudo-critical point, which again moves closer to $g_c$ as the
correlator size increases. The magnetization close to $g_{c}$ has not completely
converged for the larger plaquette correlators in our calculations: the difference in
$\langle \sigma_{x} \rangle$ for the $4 \times 4$ plaquette correlator and the $3 \times
3$ plaquette correlator is 0.003. Comparing this with the convergence found using TTNs,
for a $10 \times 10$ system, the difference in the transverse magnetization for $\sim
10^6$ parameters and $\sim 10^7$ parameters is around
$0.002$.\cite{Tagliacozzo:2009p3620} We see some scatter in the behavior of both the
absolute and transverse magnetization calculated using bond correlators, owing to the
more difficult minimization, which means that the convergence cannot be accurately
determined. However, despite this noise it is clear from our results that bond
correlators display a pseudo-critical point much closer to that found in Ref.
\onlinecite{Hamer:2000p3494} for $r_{\mathrm{max}}\ge 5$ as indicated by both the
absolute and transverse magnetization. This suggests that bond correlators are capturing
the critical behavior of the local order parameters in the TIM better. It is perhaps
surprising that the energy density is better described by plaquette correlators, while
the behavior of the local order parameter is qualitatively better with bond correlators.
We next study the performance of plaquette and bond correlators at describing the
long-range correlations, to provide some further insights.

\begin{figure}[t]
\begin{center}
\includegraphics{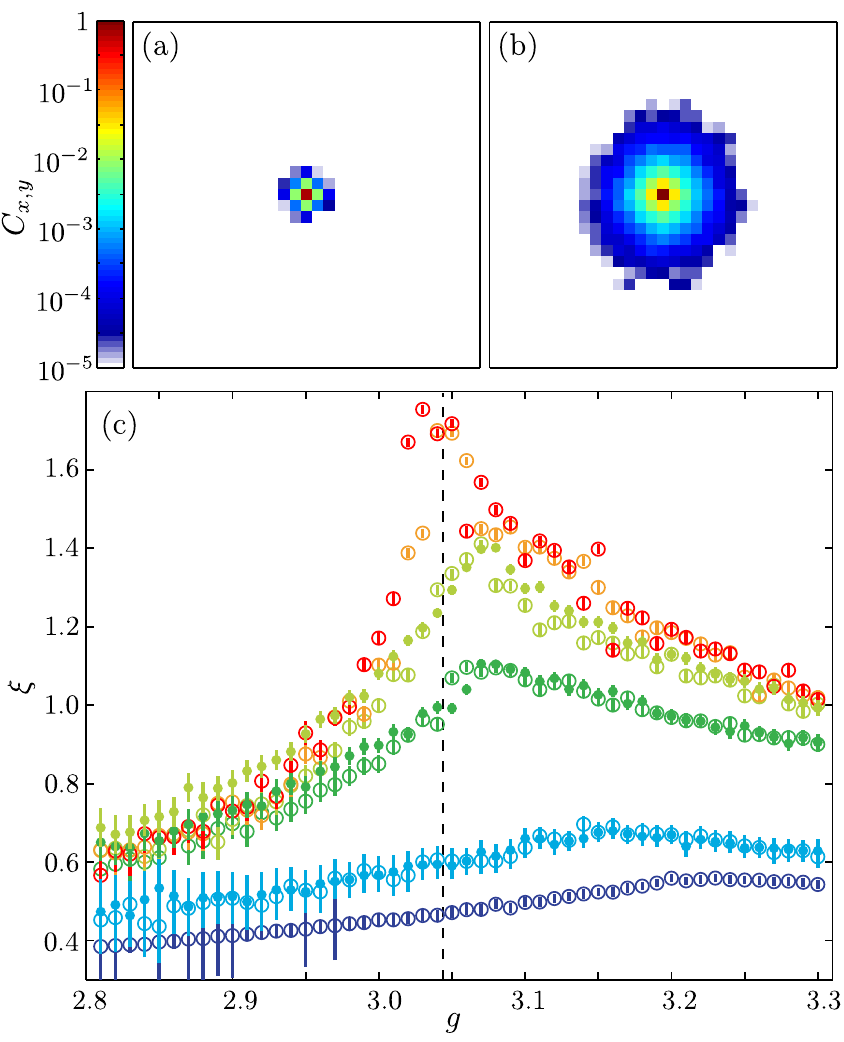}
\caption{(Color online) The connected two-point correlation function plotted using a
logarithmic color scale in a $31 \times 31$ system at $g = 3.05$ found using (a) n.\ n.\
correlators and (b) bond correlators with $r_{\mathrm{max}} = 8$. (c) Correlation length
as a function of $g$ for different correlator types, bond correlators are open circles,
plaquette correlators are closed circles. See Fig.\ \ref{fig:mvBmp} for full legends. The
dashed line indicates the critical magnetic field found by a finite-size scaling analysis
(Ref.\ \onlinecite{Hamer:2000p3494}).} \label{fig:corrising}
\end{center}
\end{figure}

\subsection{Long-range correlations}
The success of the MPS and PEPS-type tensor network approach is intimately connected with
the decay of correlations in the system. When the correlations are short range, the
tensor network representation is likely to be able to model the system well, with a
number of parameters exponentially smaller than the Hilbert space size. When the
long-range correlations approach polynomial decay, however, it can be difficult to model
the system accurately using these methods, and a larger number of parameters are required
to describe the entanglement in the system.  Conversely, the success of MERA and TTNs is
based on the logarithmic rather than linear scaling of the distance in the network
between two points a given distance
apart,\cite{MERA1,Giovannetti2008,MERAcritical,Tagliacozzo:2009p3620} and this ability to
describe long range correlations can dramatically aid performance.

To investigate how the success of the CPS description is related to the ability to
describe long range correlations we calculate the connected two point correlation
function given by
\begin{equation} \label{eq:corr}
C_{x,y} = \langle \sigma_z^{[0,0]}\sigma_z^{[x,y]} \rangle - \langle |\sigma_z^{[0,0]}|
\rangle\langle |\sigma_z^{[x,y]}| \rangle,
\end{equation}
for a wave function that has been minimized at different values of the transverse
magnetic field and for different correlator types. The connected two point correlation
function $C_{x,y}$ describes the probability of two spins on separated sites being
aligned with one another, and can only be non-zero if entanglement exists between these
two sites in the underlying ground state. When the system is close to the pseudo-critical
point, it possesses its maximum correlation length $\xi$, and the system is the hardest
to simulate numerically.

We calculate the average connected two-point correlation function in a $31 \times 31$
system. Figures \ref{fig:corrising}(a) and \ref{fig:corrising}(b) show a plot of
$C_{x,y}$ for n.\ n.\ bonds and bonds with $r_{\mathrm{max}}= 8$ respectively. These show
that the correlations grow substantially with only a moderate bond length increase. We
also determine the correlation function $C_l = C_{x,y}$ as a function of distance $l$
between the central point $(0,0)$ and the point $(x,y)$ and calculate the correlation
length $\xi$ by fitting to an exponential decay $C_l = \exp(-l/\xi)$ for different
correlator types. We plot this as a function of transverse magnetic field $g$ in Fig.\
\ref{fig:corrising}(c) and the behavior illustrates the longer reach of bonds. We see a
peak in the correlation length, consistent with the transition point seen in the
magnetization. This peak becomes sharper and moves from $g \sim 3.15$ for smaller bond
correlators and plaquette correlators to $g \sim 3.05$ for the largest bond correlators,
which is closer to the predicted value in the thermodynamic limit. However, even with the
largest bond size, the correlation length does not exceed significantly more than a
single lattice site, and thus seem to poorly model the true divergence in this property.
It is clear that using nearest-neighbor bonds to describe the system discards much of the
information about long-range correlations in the system. The increasing size of the peak
gives some indication of a slowly growing divergence, with bond correlators having
$r_{\mathrm{max}} = 8$ showing the largest correlation length.

The above results show that bond correlators are not only better for capturing the
signatures of critical behavior as reflected in the local order parameters, they are also
better at capturing the long-range correlations despite having a larger energy estimate
than plaquettes. Note that it is not uncommon for a state with a higher energy to
represent other properties of the ground state more accurately than that with the lower
energy estimate.\cite{Messiah}  For this reason it is important that the choice of ansatz
should try to reflect some expected underlying properties of the ground state and that
the results, beyond just energy, need to be carefully examined (as in our work). The
energy of a given state depends entirely on short-range correlations. Plaquette
correlators have a large number of parameters to describe the short-range interactions,
while the comparatively small number of parameters for bond correlators limit the
accuracy to which it can describe the energy. However the structure of bond correlators
allows an efficient description of any prevalent long-range correlations since it
provides direct bonds between more distant lattice sites than can be reached with
plaquette correlators. Although the calculated correlation length is still not more than
two lattice sites close to the pseudo-critical point, it is for this reason that bond
correlators are better able to describe the longer ranged properties of the Ising ground
state, as found in this investigation.

\begin{figure}[t]
\begin{center}
\includegraphics{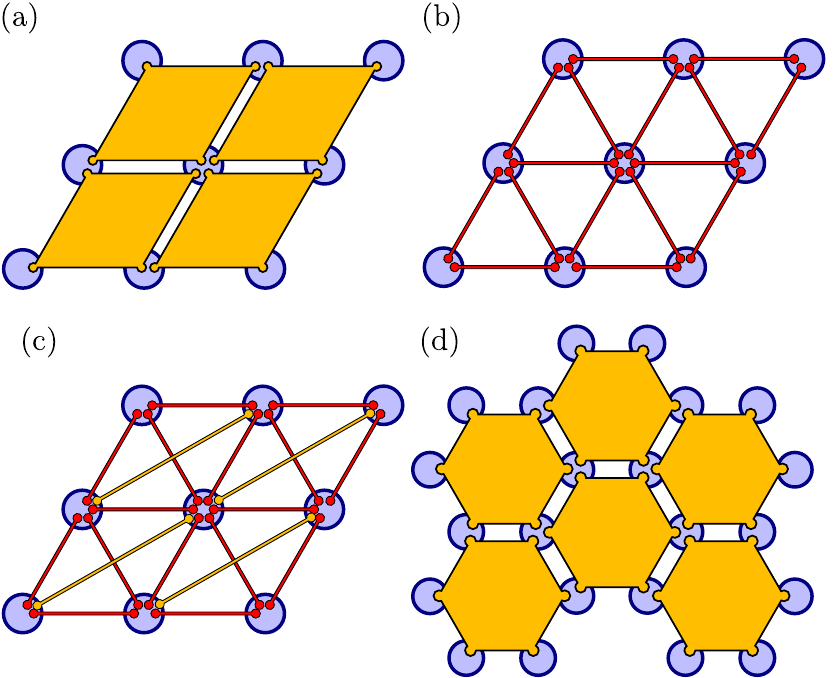}
\caption{(Color online) Correlators used in triangular and hexagonal lattice geometries.
(a) Four-site plaquette correlators on a triangular lattice. (b) n.\ n. bond correlators
on a triangular lattice. (c) $r_{\mathrm{max}} = 1$ bond correlators on a triangular
lattice. (d) Six-site plaquette correlators on a hexagonal lattice.}
\label{fig:trihexsetup}
\end{center}
\end{figure}

\section{Different geometries: Ising model on triangular and hexagonal lattices}  \label{sec:istrihex}

\begin{figure}
  \includegraphics{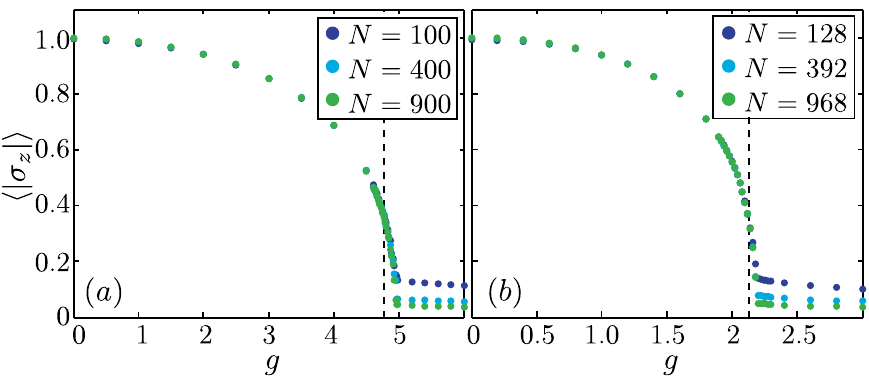}\\
  \caption{(Color online) Absolute magnetization as a function of transverse magnetic field
  for lattices with different numbers of sites $N$ in (a) a triangular lattice with the ground
  state found using four-site plaquette correlators and (b) a hexagonal lattice with the
  ground state found using six-site plaquette correlators.
   The dotted line in both figures shows the position of the critical point
calculated using a finite-size scaling analysis (Ref.\
\onlinecite{OitmaaIsing}).}\label{fig:trihexplot}
\end{figure}

To generalize the above conclusions to different lattice geometries, we also apply the
CPS ansatz and minimization method to triangular and hexagonal lattices using the
correlator types illustrated in Fig.\ \ref{fig:trihexsetup}. The behavior of the system
is expected to be qualitatively the same for the square lattice, with the critical
magnetic field shifted due to the different coordination numbers. For hexagonal and
triangular lattices, finite-size scaling analysis predicts $g_c = 2.13$ and $g_c = 4.77$
respectively.\cite{OitmaaIsing} We  calculate the absolute magnetization as a function of
transverse magnetic field for the smaller plaquette correlators, for different lattice
sizes, and the results are shown in Fig.\ \ref{fig:trihexplot}. We see the expected
behavior, with the pseudo-critical point shifted to a higher magnetic field for small
plaquette correlators, and find that the value of the absolute magnetization at $g \gg
g_c$ decreases for increasing lattice size $L$.

\begin{figure}
  \includegraphics{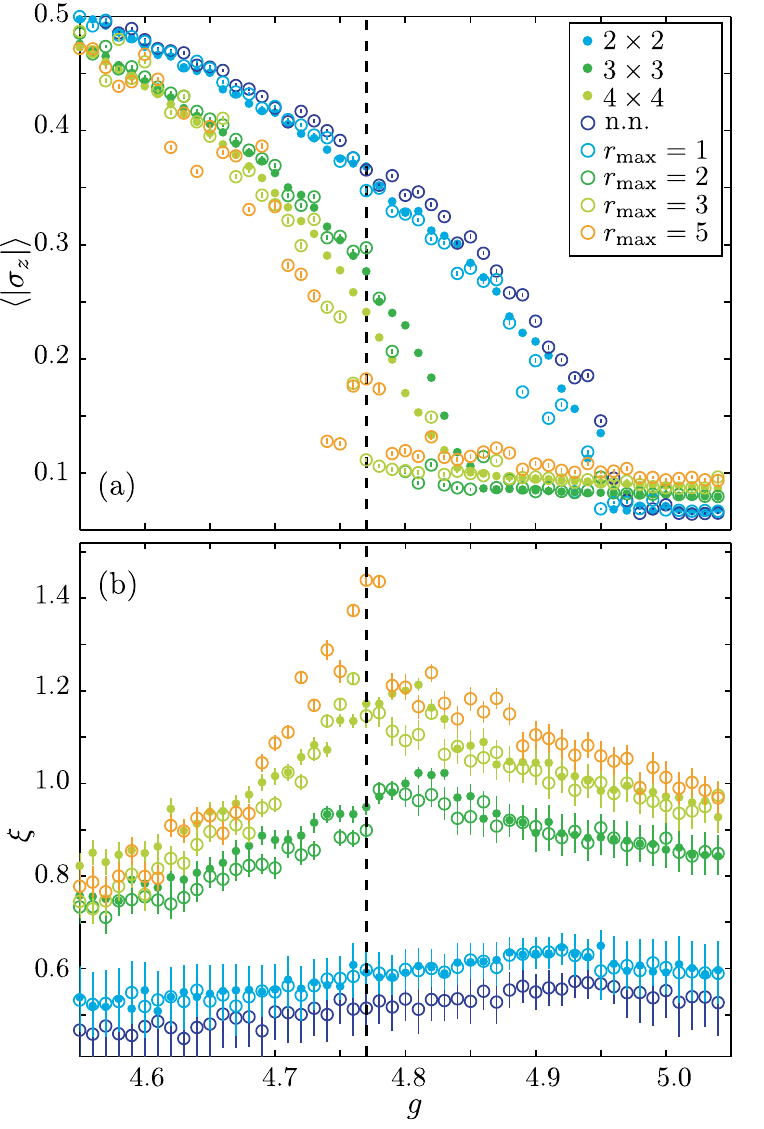}\\
  \caption{(Color online) (a) Absolute magnetization as a function of transverse magnetic
  field in a 400-site triangular lattice, for the ground
state found using different-sized plaquette correlators (closed circles) and bond
correlators (open circles). (b) Correlation length as a function of transverse magnetic
field found using different-sized plaquette correlators (closed circles) and bond
correlators (open circles). The dotted line in both figures shows the position of the
critical point calculated using a finite-size scaling analysis (Ref.\
\onlinecite{OitmaaIsing}).}\label{fig:triplot}
\end{figure}

As with the square lattice, we calculate energy close to the pseudo-critical point, and
we find that plaquette correlators give the lowest energy. For example, for a triangular
lattice with 400 sites at $g = 4.77$ the energy found using $4 \times 4$ plaquettes is
$-4.9899(1)J$, while with bonds having $r_{\mathrm{max}} = 5$ the minimum energy is found
to be $-4.9895(2)J$. Also as in square lattices we find that despite this, bond
correlators seem to be better for determining the critical behavior of the system. Figure
\ref{fig:triplot}(a) shows the absolute magnetization as a function of transverse
magnetic field for different correlator types. As with the square lattice, the longer
range of the bond correlators predict a pseudo-critical point that is closer to the
critical point (although once again more scatter is evident). The correlation length as a
function of transverse magnetic field in shown in Fig.\ \ref{fig:triplot}(b). Again, we
find that the largest correlation length for bond correlators, and the position of the
pseudo-critical point is consistent with that predicted by the local order parameter.

The numerical calculations indicate that the results found for the TIM are not dependent
on the lattice geometry chosen. For all the geometries investigated it is found that the
longer ranged bond correlators predict a critical point closer to that calculated in
previous works, and can better describe the long-range correlations that are expected
close to criticality, despite the lower minimized energy found using plaquette
correlators.

\section{Antiferromagnetic order: the Heisenberg model} \label{sec:heis}
To examine how plaquette and bond correlators compare when describing very different
physics, we also investigate the AFHM on a square lattice. This system is more
challenging to describe than the TIM, as it exhibits richer behavior owing to its
symmetry and antiferromagnetism. The antiferromagnetic spin-$\frac{1}{2}$ Heisenberg
model is described by the Hamiltonian
\begin{equation}
H_0 =
J\sum_{i=1,j=1}^{L}\mathbf{S}^{[i,j]}\cdot\mathbf{S}^{[i+1,j]}+\mathbf{S}^{[i,j]}\cdot\mathbf{S}^{[i,j+1]},
\end{equation}
where $\mathbf{S} = \frac{1}{2}(\sigma_x,\sigma_y,\sigma_z)$ is the spin-$\frac{1}{2}$
operator. It corresponds to the half-filled band limit of the Hubbard model, and is
assumed to describe the antiferromagnetic undoped insulator La$_2$CuO$_4$ as well as
other undoped copper oxide materials.\cite{Heisreview}

For a pair of spins, the ground state of $H_0$ is simply a spin singlet. However, due to
the monogamy of entanglement, each spin cannot be in a singlet state with more than one
neighbor, and the ground state for $N$ spins has to satisfy both local minimization and
global translational symmetries giving a highly complicated superposition
state.\cite{Verstraete:2008p3835} As a result of this, it is predicted that long-range
order arises.\cite{PhysRevB.39.2344, *PhysRevB.49.11919} This is characterized by the
two-point correlation function for the total spin $ \langle
\mathbf{S}^{[a]}\cdot\mathbf{S}^{[b]}\rangle $ and the staggered magnetization, which is
reduced from the N\'{e}el, or classical, value by quantum fluctuations.\cite{Heisreview}

The AFHM has also already been treated using plaquette type
correlators,\cite{changlani2009,Mezzacapo2009} but here we extend that description by
comparing them with bond correlators, and examine which correlator type better captures
the expected long-range order. The results obtained here are also compared with those
found using SSE, which give the current best estimate.\cite{PhysRevB.56.11678} This
system has been treated using a resonating-valence bond (RVB) picture, by considering
states that are superpositions of all possible coverings of singlets on the
sites.\cite{RVBHeis, SandvikVB, lousvk2007,SandvikEvertz} The latest
study\cite{SandvikEvertz} gives estimates of the energy and staggered magnetization that
agree well with those found using SSE.

We restrict to configurations with an equal number of up and down spins, and move between
configurations by flipping a pair of opposing spins (so that the net magnetization is
unchanged). This can be done since the ground state exists in the spin sector having
$\sum_{i,j}\sigma_z^{[i,j]} = 0$.\cite{MarshallRule} We assume real, positive correlators
and use the Marshall-sign rule to determine the correct sign for each
configuration.\cite{MarshallRule} Otherwise we apply the same minimization routine as
used above for the TIM. The ground state energy was found for different plaquette types
and lattice sizes, and the results for a $14 \times 14$ lattice are shown in Table
\ref{tb:Heisen}.

\begin{table}
\begin{center}
\begin{tabular}{|c||c|c|}
  \hline
  Correlator type & No. of elements. & Energy  \\
  \hline
  \hline
  ${r}_{\mathrm{max}} = 1$ & 16 & $-0.6560(3)$\\
  ${r}_{\mathrm{max}} = 2$ & 48 & $-0.6627(3)$\\
  ${r}_{\mathrm{max}} = 3$ & 96 & $-0.6641(2)$\\
  ${r}_{\mathrm{max}} = 4$ & 160 & $-0.6646(1)$\\
  ${r}_{\mathrm{max}} = 5$ & 240 & $-0.6648(3)$\\
  ${r}_{\mathrm{max}} = 6$ & 336 & $-0.6647(2)$\\
  ${r}_{\mathrm{max}} = 7$ & 504 & $-0.6649(3)$\\
  \hline
  $2 \times 2$ plaquettes & 16 &  $-0.6567(3)$\\
  $3 \times 3$ plaquettes & 512 & $-0.6662(2)$\\
  $4 \times 4$ plaquettes & 65,536 & $-0.6685(4)$\\
  \hline
\end{tabular}
\end{center}
\caption{Energy per site $E/J$ found using stochastic minimization for the AFHM in a $14
\times 14$ system for different correlator types. The result from using a SSE is $E/J =
-0.670222(7)$ (Ref.\ \onlinecite{PhysRevB.56.11678}).} \label{tb:Heisen}
\end{table}

For plaquette correlators, the energies obtained are consistent with previous
calculations.\cite{changlani2009,Mezzacapo2009} We see that plaquette correlators are
able to minimize the energy much better than bond correlators, and that the discrepancy
between the correlator types is larger than with the TIM. For example, for $L=14$, even
$3 \times 3$ correlators are better able to minimize the ground state energy than bonds
with $r_\mathrm{max} = 7$, where bonds are long enough to couple every pair of sites on
the lattice, and where the number of parameters is comparable. Thus the relative behavior
of bond correlators with respect to plaquette correlators is worse than with the TIM.
This is likely to be as a result of a more complicated energy landscape for the AFHM than
for the TIM, such that the fragmented structure imposed by the bond correlators makes it
difficult to describe the ground state. Consistent with the RVB picture, it therefore
appears that having a large number of local parameters is more important to accurately
describe the ground state of this model than to provide longer-ranged parameters. This is
further highlighted by the fact that the energy estimated using bond correlators with
$r_\mathrm{max} = 4$ is the same (within error) as the energy estimated with
$r_\mathrm{max} = 7$, so the improvement with increasing bond size is seen to saturate.

As with the TIM, we also investigate how well the correlations are described by the
different correlator types by calculating the staggered magnetization. The staggered
magnetization has been previously calculated using a variety of tensor network
methods,\cite{TRG1,PhysRevB.78.205116,iPepsHeis,PhysRevB.81.174411} including correlator
product states.\cite{Mezzacapo2009} It has been found to be challenging to determine
accurately, with the estimated value in general being larger than the value found using
SSE.\cite{PhysRevB.56.11678} The staggered magnetization $M_1(L)$ can be defined using
the two-point correlation function at the furthest point out in the lattice:
\cite{Reger1998}
\begin{equation}
M_1^2(L) = C_{L/2,L/2} = \langle \mathbf{S}^{[0,0]}\cdot\mathbf{S}^{[L/2,L/2]}\rangle.
\end{equation}
An alternative definition for the staggered magnetization is:  \cite{Reger1998}
\begin{equation}
M_2^2(L) = \frac{1}{L^2}\sum_{i,j}(-1)^{i+j}C_{i,j}.
\end{equation}
These scale to the (same) staggered magnetization in the thermodynamic limit $M(\infty)$.
The scaling behavior of $M_{1,2}$ has been predicted using chiral perturbation theory and
spin-wave theory, and is given by $M_{1,2}^2(L) = M(\infty)^2 + b_{1,2}/L +
\cdots$.\cite{PhysRevB.56.11678,Hasenfratz93,Huse88}

\begin{figure}[t]
\begin{center}
\includegraphics{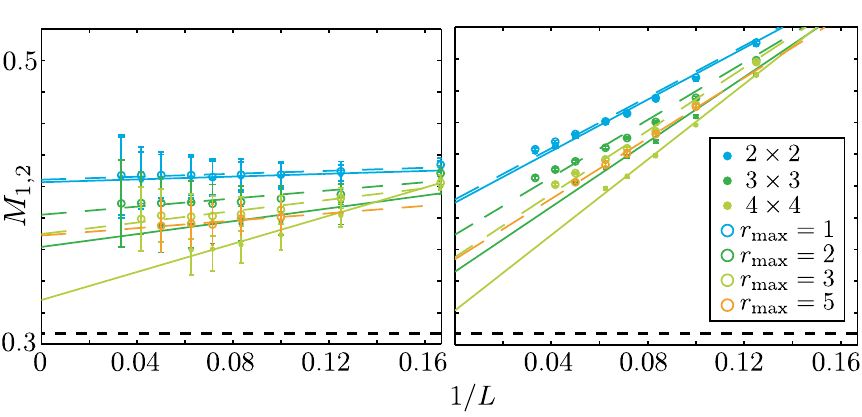}
\caption{(Color online) Staggered magnetization found using different plaquette types
against the inverse of the lattice size $L$. (a) shows the staggered magnetization $M_1^2
= C_{L/2,L/2}$ and (b) shows the staggered magnetization $M_2^2(L) =
\frac{1}{L^2}\sum_{i,j}(-1)^{i+j}C_{i,j}$. In both plots the black dashed line denotes
the staggered magnetization in the thermodynamic limit found using SSE (Ref.\
\onlinecite{PhysRevB.56.11678}).} \label{fig:mvL}
\end{center}
\end{figure}

We calculate the two-point correlation function, and similar to a previous
work,\cite{Wang:2010p7758} it is found that the $z$-component of the correlations departs
from the $x,y$-component of the correlations --- the $z$ correlations tend to zero while
the $x,y$ correlations remain finite. This is likely to be due to the preferential
treatment of the $z$ axis in the Monte Carlo method and inherent in the CPS correlator
elements, i.e.\ the chosen basis. We also find that very small changes in the estimated
ground state energy can lead to large differences in the estimated staggered
magnetization. The results of these calculations are shown in Fig.\ \ref{fig:mvL}. In
contrast to the results in previous sections where bond correlators performed better, we
find that plaquette correlators provide a better estimate of the long-range order, with
an extrapolated value of $M_1(\infty)=0.33(1)$ and $M_2(\infty)= 0.322(7)$ found using $4
\times 4$ plaquette correlators closest to the value $M(\infty) = 0.3070(3)$ found using
stochastic series expansion.\cite{PhysRevB.56.11678} The value found in a previous study
using the RVB basis is $M(\infty) = 0.30743(1)$.\cite{SandvikEvertz} Unlike with the TIM,
we find that it is not the range of the correlators that determines the accuracy with
which correlations can be represented, but instead seems to be the number of local
parameters. Note that although the RVB model has a similar pairwise structure to bond
correlators, the description used here is not equivalent to the RVB model, which does
seem better able to represent the ground state. This is most likely due to the singlet
structure of the RVB model, which can preserve the symmetry between the three spin axes
exactly. The relative behavior of bond and plaquette correlators is consistent with the
minimized energy, and highlights the importance in the Heisenberg model of describing
accurately the more complicated local correlations between multiple nearby spins in the
three spin axes.

\section{Summary and outlook} \label{sec:conc}
We have investigated the performance of the most common forms of correlator product
states for describing the ground state of the TIM in various two-dimensional geometries.
Even correlators with a small number of parameters (16 parameters for the $2 \times 2$
plaquette correlator) can describe the qualitative behavior of large systems (up to 51
$\times$ 51 spins). Increasing the correlator size to $4 \times 4$ site plaquettes or
bonds with $r_{\mathrm{max}} \ge 3$ provides accurate results for up to $L=31$ (and in
principle larger systems could be studied). We have also found that, despite having a
slightly larger minimized energy, bond correlators are more successful than plaquette
correlators at describing critical behavior, such as the long-ranged correlations, whilst
using a smaller number of parameters and having a faster computation time. We have also
investigated the AFHM in a square lattice. Results with plaquette correlators are
comparable with previous works,\cite{Mezzacapo2009} and we have found that, unlike the
TIM, plaquette correlators are better able to describe both the energy and the long range
behavior of the ground state.

The system sizes reached here are comparable with experimental cold-atom lattice systems,
and the ability to efficiently model spin systems is useful for rapidly progressing
cold-atom simulations of quantum magnetism.\cite{QMag1,QMag2} A general result for the
optimal correlator structure for a given system is not straightforward to deduce, but our
findings still provide useful insights for applications to condensed matter systems. For
example, we would expect plaquette correlators to be the optimal choice for the $J_1-J_2$
model studied in Ref. \onlinecite{Mezzacapo2009} and for the Fermi-Hubbard model.
However, we may expect bond correlators to perform better for the Bose-Hubbard model,
where the choice of the local Fock basis may allow a good description of the ground state
with a smaller number of local parameters, or in spin-systems which are highly
anisotropic like the TIM. In addition, the Laughlin wave function\cite{LaughlinPRL} has
an exact CPS description using bond correlators.\cite{changlani2009}

It may be that an optimal CPS ansatz consists of a hybrid of bond and plaquette
correlators, such as strings of sites (in a similar set up to string bond states) linking
all sites on the lattice, or extended rectangular plaquette correlators. This would
provide a larger number of local parameters while still allowing the correlators to have
a long range. In further work,\cite{Inprep} the CPS description will be applied to other
systems, such as the Bose-Hubbard model in the presence of an artificial magnetic field,
\cite{2008AdPhy..57..539C,*Fetter:2009p6403,*Dalibard:2010p3588,*Williams:Vortices} in
the regime where the fractional quantum Hall states are predicted to
occur.\cite{Sorensen:2005p2036,*Palmer:2006p47,*Moller:2009p3050}

\section{Acknowledgements}
SA thanks Vlatko Vedral and Nigel Cooper for useful discussions. This work was supported
by the UK EPSRC through project EP/J010529/1 and by the ESF program EuroQUAM (EPSRC grant
EP/E041612/1). SRC and DJ thank the National Research Foundation and the Ministry of
Education of Singapore for support.

\appendix

\section{Exact and efficiently samplable tensor network states} \label{ap:samp}
For the case of n.\ n.\ bond correlators, the resulting CPS is in fact equivalent to an
MPS with internal dimension $\chi$ equal to the physical dimension
$d$.\cite{changlani2009} Note that this equivalence does not hold in the reverse
direction, i.e.\ not all MPS with $\chi=d$ are equivalent to a CPS. Correlator product
states can be thought of as a special class of matrix product states that can be
factorized using the copy tensor (also known as the diagonal in category theory, the
COPY-gate or the COPY-dot in circuits).\cite{CTNS} The copy tensor copies a chosen set of
physical basis states e.g.\ $|x\rangle$ where $x = \uparrow, \downarrow$, and
subsequently breaks up into disconnected states.  This is essential for such states to be
efficiently and exactly sampled.

\begin{figure}[htb]
\begin{center}
\includegraphics{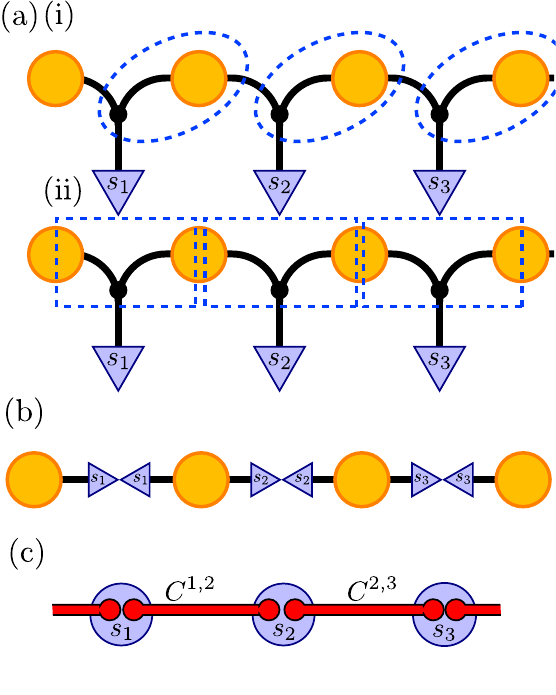}
\caption{(Color online) Illustration of the equivalence between matrix product states and
correlator product states for the simple case of bond correlators, when the dimension of
the correlators is the same as the physical dimension: (a) Shows the correlators
connected by copy tensors (black dots), which is equivalent to an MPS where (i) each MPS
tensor (dashed ellipse) is formed by the copy tensor and the neighboring correlator or
(ii) each correlator is `split' (e.g. using a singular value decomposition) and each MPS
tensor (dashed box) is formed from a copy tensor and the adjacent correlator factors; (b)
this can be factorized, breaking up the connections while ensuring the same configuration
is present on neighboring correlators; (c) this gives a product of scalar correlator
elements.} \label{fig:cpsmps}
\end{center}
\end{figure}

This is illustrated in Fig.\ \ref{fig:cpsmps}(a), where the copy tensor is represented
using a black dot, while the large circle represents a matrix formed of the correlator
elements. The copy tensor has one input (the physical leg) and multiple outputs (two for
the case of bond correlators in one dimension). This network is equivalent to an MPS,
where each matrix in the network can be formed from the correlators and copy tensors. The
copy tensor acts by taking the value of the physical leg in the chosen basis and copying
it to all its legs. As shown in Fig.\ \ref{fig:cpsmps}(b) this acts to break up the bonds
between the matrices while ensuring that neighboring matrices are contracted according to
the same physical index on a pair of virtual legs. The result of this is a product of
scalars, shown in Fig.\ \ref{fig:cpsmps}(c), which are the correlators for the given
input configuration used.

\begin{figure}
  \includegraphics{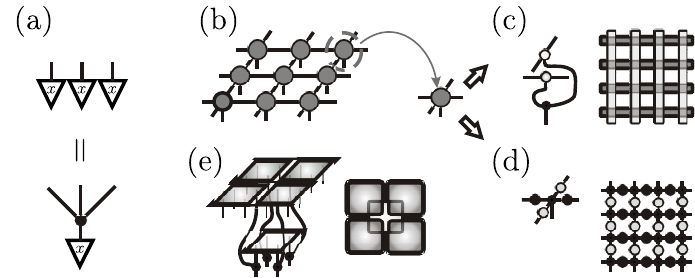}\\
  \caption{(a) The copy tensor copies computational basis states and subsequently breaks up into disconnected states. (b) A
generic PEPS in which we expose a single generic rank-5 tensor. This tensor network can
neither be contracted nor sampled exactly and efficiently. However, if the tensor has
internal structure exploiting the copy tensor then efficient sampling becomes possible.
(c) The tensor breaks up into a vertical and a horizontal rank-3 tensor joined by the
COPY-dot. Upon sampling computational basis states the resulting contraction reduces to
many isolated MPS, each of which are exactly contractible, for each row and column of the
lattice. This type of state is known as a string-bond state and can be readily
generalized (Ref.\ \onlinecite{Schuch2008sbs}). (d) An even simpler case is to break the
tensor up into four rank-2 tensors joined by a copy tensor forming a nearest-neighbor
bond correlator-product state. (e) Plaquette correlators, which are outside the PEPS
class, join up overlapping tensors (in this case rank-4 ones describing a $2 \times 2$
plaquette) for each plaquette. Efficient sampling is again possible due to the copy
tensor.}\label{fig:ctns}
\end{figure}

In two dimensions, correlator product states can also be considered to be equivalent to a
special case of PEPS --- in this case the copy tensor copies the physical index to four
different correlator matrices. String bond states are another special factorization of
PEPS where the contractible components are themselves MPS, rather than a correlator
matrix.\cite{Schuch2008sbs} States such as these can then be sampled exactly-efficiently,
as they consist of product of scalar values, each scalar value only depending on a small
number of physical indices. This is illustrated in Fig.\ \ref{fig:ctns}.

Since the MPS representation is exhaustive, every correlator product state has an
equivalent matrix product state. For certain systems, however, it may be far more
efficient to use the CPS representation. For example, in the MPS representation
long-range correlations are mediated by the bonds between neighboring sites. Thus, in
general, a very large bond-dimension may be needed between neighboring tensors to capture
the entanglement between sites spaced by a large distance. However, using CPS it is
possible to include a correlator directly between distant sites, describing the
entanglement with a small bond dimension since the bond directly links the two sites.
Correlator product states can thus describe systems that do not obey an area law.

\section{Numerical methods} \label{ap:meths}

\subsection{Calculating expectation values} \label{ap:expvals}
Expectation values of the energy and other operators are determined from a CPS by using
Monte Carlo importance sampling. The energy $E$ is given by
\begin{equation}
E = \langle \psi |H|\psi \rangle  =
\frac{\sum_{\mathbf{s},\mathbf{s}'}W^{\ast}(\mathbf{s}')\langle\mathbf{s'}|H|\mathbf{s}\rangle
{W(\mathbf{s})}}{\sum_{\mathbf{s}}|W(\mathbf{s})|^{2}},
\end{equation}
where we assume that the wave function is not normalized, as is generally the case when
using CPS. The energy can be expressed as a weighted sum $E =
\sum_{\mathbf{s}}P(\mathbf{s})E(\mathbf{s})$ which is as a sum over configurations of the
product of the local energy $E(\mathbf{s})$ and the probability $P(\mathbf{s})$, given by

\begin{equation} \label{eq:prob}
E(\mathbf{s}) =
\sum_{\mathbf{s}'}\frac{W^{\ast}(\mathbf{s}')}{W^{\ast}(\mathbf{s})}\langle\mathbf{s'}|H|\mathbf{s}\rangle,P(\mathbf{s})
= \frac{|W(\mathbf{s})|^{2}}{\sum_{\mathbf{s}}{|W(\mathbf{s})|^{2}}}.
\end{equation}

By replacing the Hamiltonian $H$ with any operator $O$ its expectation value $\langle
\psi|O|\psi\rangle$ can be similarly computed. For example, for the TIM described in
section \ref{sec:isres}, calculation of the interaction energy terms
$\sigma^{[i,j]}_{z}\sigma^{[i+1,j]}_{z}$ is straightforward since the operators are
diagonal: the local energy $E_{zz}(\mathbf{s})$ is given by
\begin{equation}
E_{zz}(\mathbf{s}) =-\sum_{i=1,j=1}^{L}(s^{[i,j]}s^{[i+1,j]}+s^{[i,j]}s^{[i,j+1]}),
\end{equation}
where $s^{[i,j]} = \pm1$. In a translationally invariant system, the sum over all sites
can be dispensed with and the exchange energy simply taken with respect to the first (or
any other) spin and its neighbors.

To calculate the transverse energy terms $g\sigma_{x}^{[i]}$, we use $\sigma_{x} =
\sigma_{+} + \sigma_{-} $. The only non-zero terms of the energy estimator
$E(\mathbf{s})=\sum_{\mathbf{s}'}\frac{W(\mathbf{s}')}{W(\mathbf{s})}\langle\mathbf{s'}|H|\mathbf{s}\rangle$,
are those for which $\mathbf{s}'$ differs from $\mathbf{s}$ by a single spin-flip, so the
local transverse energy is given by
\begin{equation}
E_{x}(\mathbf{s}) =-\sum_{i=1,j=1}^{L}\frac{W(\mathbf{s}'_{ij})}{W(\mathbf{s})}.
\end{equation}
where $\mathbf{s}'_{ij}$ denotes that the spin on site $(i,j)$ has been
flipped.\cite{SandvickVidal2007} The absolute magnetization $\langle |\sigma_{z}
|\rangle$ of the ground state can also be calculated. Note that the expectation value of
the magnetization $\langle \sigma_{z} \rangle$ is always be zero, since the global
$\mathbb{Z}_2$ symmetry of the system means the ground state forms equal superpositions
of configurations with all spins flipped. However, the absolute magnetization quantifies
how well the spins are aligned with one another, and is found by summing the local
absolute magnetization
\begin{equation}
|\sigma_{z}(\mathbf{s})| = \frac{1}{L^{2}}\left|\sum_{i=1,j=1}^{L} s^{[i,j]} \right|.
\end{equation}

The probability of a given configuration is never explicitly calculated using equation
(\ref{eq:prob}). Instead, the configurations are visited according to importance
sampling, using the Metropolis Algorithm.\cite{Metropolis:1953p6006} New configurations
$\mathbf{s}'$ are generated by randomly flipping spins according to the calculated
acceptance probability $|W(\mathbf{s}')/W(\mathbf{s})|^2$. There are a total of $F\times
N$ spin flips per bin, where we call $F$ the number of sweeps per bin, and $N$ is the
number of spins (since we visit each spin sequentially in a given sweep). As is typical,
we also first perform a few warm-up sweeps that do not contribute to the estimates to
ensure that the random walk through configuration space starts in an equilibrium
position, and also sample the local expectation value only every $\texttt{s}^{th}$ sweep,
where $\texttt{s}$ is the sample rate $\sim 10$, to allow a larger portion of the
configuration space to be visited and to ensure that the configurations that are sampled
are independent from one another.

At no point during the calculation is knowledge of the normalization of the wave function
needed. The most time-consuming step in the calculation is determining the correlator
fraction $W(\mathbf{s}')/W(\mathbf{s})$. Since the  configuration weight is given by a
simple product of numbers, to calculate this fraction only the correlators that represent
the sites where the configurations have changed are required. If the Hamiltonian is local
then there are only a few configurations $\mathbf{s}'$ that have a non-zero matrix
element in the local energy, so only a few correlator fraction terms need to be
calculated for each contribution to the energy.

\subsection{Energy minimization} \label{ap:emin}

The correlator elements that minimize the energy are found using a stochastic
minimization method which requires only the sign of the first derivative of the energy
with respect to the correlator elements.\cite{lousvk2007} Specifically, the first
derivative is given by
\begin{equation}
\frac{\partial E}{\partial C^{\{i\}}_{s_{\{i\}}}} = 2(\langle \Delta^{\{i\}}_{s_{\{i\}}}
E \rangle - \langle \Delta^{\{i\}}_{s_{\{i\}}}\rangle\langle E \rangle), \label{eqn:dvt}
\end{equation}
where $\Delta^{\{i\}}_{s_{\{i\}}}(\mathbf{s}) $ is given by
\begin{equation}
 \Delta^{\{i\}}_{s_{\{i\}}}(\mathbf{s}) = \frac{1}{W(\mathbf{s})}\frac{\partial W(\mathbf{s})}{\partial C^{\{i\}}_{s_{\{i\}}}}
 \end{equation}
This is trivial to compute, since $W(\mathbf{s})$ is simply a product of the different
correlators $C$. If each correlator is only used once
$\Delta^{\{i\}}_{s_{\{i\}}}(\mathbf{s}) = 1/C^{\{i\}}_{s_{\{i\}}}$, whereas if the same
correlator is used for multiple sites (e.g. in a translationally invariant system) then
$\Delta^{\{i\}}_{s_{\{i\}}}(\mathbf{s}) = b^{\{i\}}/C^{\{i\}}_{s_{\{i\}}}$, where
$b^{\{i\}}$ is the number of times the correlator $C^{\{i\}}_{s_{\{i\}}}$ appears in the
product for the correlator amplitude for configuration $\mathbf{s}$.

Following the procedure in Ref. \onlinecite{SandvickVidal2007}, the expectation value of
the derivative  is calculated in the same way that other operators are calculated for $F$
sweeps in a bin, and the derivatives for all correlators are calculated simultaneously.
After this every correlator is updated according to
\begin{equation}
C^{\{i\}}_{s_{\{i\}}} \rightarrow C^{\{i\}}_{s_{\{i\}}} - \texttt{r} \delta(k)
\mathrm{sign}\left(\frac{\partial E}{\partial C^{\{i\}}_{s_{\{i\}}}}\right)^{\ast},
 \end{equation}
where $\texttt{r}$ is a random number between 0 and 1, and $\delta(k)$ is the step-size
for a given iteration $k$. Unlike the Newton method, the second derivative is not
required, considerably simplifying the calculation. This method has been found to give
better convergence than the Newton method, since it avoids the sizable statistical errors
in the second derivative and also because it does not exactly follow the direction of
steepest decent which can be more efficient.\cite{lousvk2007}

To achieve convergence, $F$, the number of sweeps in a bin is increased every iteration,
and the step size $\delta(k)$ is reduced. For every iteration $k$, $F$ is given by
$F_{0}k$, and the number of bins is increased as $G = G_{0}k$ per iteration to slow the
cooling rate, where typically $F_{0}$ and $G_{0}$ $\sim~10$. The step size is reduced per
iteration as $\delta = \delta_{0}k^{-\eta}$. We find best results with $\eta$ between
0.75 and 0.9. After an initial run with a relatively large step size to get close to the
minimum, the resulting correlators are then used as a starting point for a new run of
iterations (i.e.\ $k$ is reset) where $F_{0}$ and $G_{0}$ remain unchanged, but
$\delta_{0}$ is decreased. Depending on the system parameters this can reduce the energy
further, while for other regimes (i.e.\ away from criticality) the energy has already
converged after the first run. After the minimization is complete, the procedure is
repeated for a single iteration with zero step size and large $F$ and $G$, to obtain an
accurate estimate of the expectation values.

\begin{figure}[t]
  \includegraphics{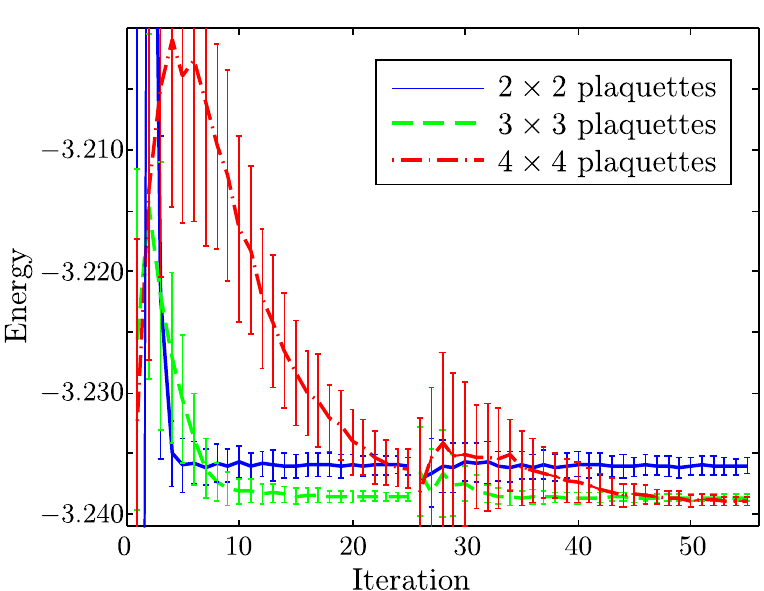}\\
  \caption{(Color online) Energy during minimization of the plaquette correlators for the two-dimensional TIM in a 31 by 31 system with g = 3.05 (see section \ref{sec:isres}). The
  initial minimization routine is performed with $\delta = 0.02$ for 25 iterations, followed by a further round of minimization with $\delta_0 = 0.005$ for 30 iterations (although not shown, for $4 \times 4$ plaquette correlators the further round of minimization has 40 iterations) .}\label{fig:minfig}
\end{figure}

For bond correlators with $r_{\mathrm{max}} < 3$ and for $2 \times 2$ plaquette
correlators, we start with a uniform state, and perform initial minimization with $F_{0}
= 10$, $G_{0} = 10$, $\delta_{0} = 0.02$, $\eta = 0.9$ for 25 iterations. For $3 \times
3$  and $4 \times 4$ plaquette correlators, the initial correlator is built using the $2
\times 2$ plaquette correlator that results from the first round of minimization.  For
bond correlators with $r_{\mathrm{max}} \geq 3$, we build approximate new correlators
using the shorter range correlator elements distributed among all bond lengths. For all
correlator types, the iteration counter is then reset, and the minimization procedure is
performed again using the minimized correlator as a starting point, and $\delta_{0} =
0.005$ for 30--40 iterations (until convergence is obtained). Accurate estimations of the
expectation values are then obtained by performing a final calculation with one iteration
having a much larger number of sweeps per bin (for example $F = 10,000$, $G = 100$,
$\delta = 0$). A typical minimization routine for plaquette correlators is shown in Fig.\
\ref{fig:minfig}.

For bond correlators, the best results are found when the step size $\delta$ is a
function of bond length as well as iteration number $k$. For each bond in the correlator,
we define the radius $r$ as the maximum displacement along one of the lattice indices,
and scale the step size as $\delta(r,k) = \delta(k)r^{-0.75}$. For the first round of
minimization we minimize the bonds for a given radius separately in sequence, while in
the second round of iterations all correlator elements are updated simultaneously.

\end{document}